# Enhanced coherent terahertz emission from critical superconducting fluctuations in $YBa_2Cu_3O_{6.6}$

D. Nicoletti[1], M. Rosenberg[1], M. Buzzi[1], M. Fechner[1], Y. Liu[2], S. Nakata[2], B. Keimer[2], R. A. Vitalone[3], D. N. Basov[3], P. E. Dolgirev[4], E. Demler[5], M. H. Michael[1], and A. Cavalleri[1,6]

*[1] Max Planck Institute for the Structure and Dynamics of Matter, 22761 Hamburg, Germany*

*[2] Max Planck Institute for Solid State Research, 70569 Stuttgart, Germany*

*[3] Department of Physics, Columbia University, New York, NY 10027, USA*

*[4] Department of Physics, Harvard University, Cambridge, Massachusetts 02138, USA*

*[5] Institute for Theoretical Physics, ETH Zurich, 8093 Zurich, Switzerland*

*[6] Department of Physics, Clarendon Laboratory, University of Oxford, Oxford OX1 3PU, United Kingdom*

## Abstract

**Coherent terahertz (THz) emission is emerging as a powerful new tool to probe symmetry breakings in quantum materials. This method relies on second order optical nonlinearities and is complementary to second harmonic generation spectroscopy. Here, we report coherent THz emission from Josephson plasmons in underdoped $YBa_2Cu_3O_{6+x}$, and find that the amplitude of the emitted field increases dramatically close to the superconducting transition temperature, $T_C$. We show theoretically how emission is enhanced by critical superconducting fluctuations, a nonlinear analogue of critical opalescence. This observation is expected to be of general importance for the study of many thermal and quantum phase transitions.**

## I. Introduction

Coherent Terahertz Emission (hereinafter referred to as CTE) spectroscopy is emerging as a powerful technique to probe subtle forms of symmetry breaking in complex solids [1,2]. Generally, CTE results either from optical rectification in non-centrosymmetric materials or from time dependent charge currents [3,4,5,6,7]. Several studies have demonstrated THz emission from complex materials, including colossal magnetoresistance manganites [6,8,9], magnetic compounds and multiferroics [10,11,12,13,14,15,16,17,18,19].

In high-$T_C$ superconductors, THz emission has been almost solely associated with time-dependent supercurrents generated by an external bias or magnetic field [7]. Amongst the various observations are near-single-cycle THz pulses in biased antennas fabricated from $YBa_2Cu_3O_{7-\delta}$ or $Bi_2Sr_2CaCu_2O_{8+\delta}$ films [4,20,21], multi-cycle narrowband emission governed by the Josephson effect in the presence of an applied out-of-plane magnetic field in $Tl_2Ba_2CaCu_2O_{8+\delta}$ films [22], and tuneable, highly efficient narrowband emission from MESA-type resonant structures made of Josephson junction stacks [23,24,25].

Recently, anomalous CTE was discovered in high-$T_C$ cuprates of the $La_{2-x}Ba_xCuO_4$ (LBCO) family, in absence of external magnetic fields and current biases [26]. This effect, which is forbidden by symmetry in this class of centrosymmetric materials, was detected only when superconductivity coexisted with *charge-stripe order* in the Cu-O planes [27], and when stripes [28,29,30] were either *incommensurate* with the lattice or *fluctuating*. These results were interpreted by reasoning that incommensurate (or fluctuating) charge density waves (CDW) [31,32] break inversion symmetry [32], and enable the emission by *surface Josephson plasmons* [33,34,35], finite momentum modes that are coupled into free space by the stripes.

In the present work, we show how charge-density-wave activated CTE exhibits striking anomalies when critical fluctuations of the order parameter set in close to $T_C$. This effect, which opens up a number of opportunities in the study of quantum phase transitions, is particularly clear in the case of $YBa_2Cu_3O_{6+x}$ (YBCO), in which the CDW survives across the superconducting phase transition, and could not be observed in LBCO, where $T_C$ and $T_{CDW}$ coincide.

$YBa_2Cu_3O_{6+x}$ has two Cu-O layers per unit cell and far higher superconducting critical temperatures than LBCO, up to ∼90 K. As in LBCO and most cuprate families, YBCO is characterized by a dynamic charge ordering in the underdoped region of its phase diagram [36,37,38,39]. Fluctuating CDWs are found in compounds near $YBa_2Cu_3O_{6.6}$ (YBCO 6.6), corresponding to a hole concentration in the Cu-O planes close to 12.5%, vanishing rapidly for doping levels away from this value [39]. In YBCO 6.6 this charge-density-wave phase appears for temperatures as high as $T_{CDW} \simeq 150$ K, more than a factor of 2 higher than the superconducting $T_C$ (see phase diagram in Fig. 1b,1c), displaying a maximum in intensity exactly at $T_C$ [39]. For lower temperatures, deep in the superconducting phase, the CDW gradually decreases and can be revived by the application of external magnetic fields, providing confirmation that it competes with superconductivity [40].

In contrast to LBCO, the CDW in YBCO is always fluctuating, as well as incommensurate with the crystal lattice (in-plane wave vector $q \simeq 0.33$). Moreover, the values of the CDW correlation length in YBCO, $\xi_{ab}\sim 50$ Å [39,40], are similar to those found in LBCO 9.5% [27], the compound for which optimal conditions for THz emission were found.

For these reasons, we investigated the THz emission properties of $YBa_2Cu_3O_{6.6}$ ($T_C = 63$ K). These were compared with the response of $YBa_2Cu_3O_{6.48}$ (YBCO 6.48), with hole concentration ~9% and $T_C = 51$ K, a compound for which only a very weak CDW below $T_{CDW} \simeq 100$ K was reported [39].

## II. Experimental results

The experimental geometry is shown schematically in the insets of Figs. 1e,1f. We used the output of an amplified Ti:Sa femtosecond laser as pump pulses, with a duration of 100 fs and photon energy of 1.55 eV (800 nm wavelength). These were focused at normal incidence onto an $ac$-cut sample surface, with polarization oriented along the out-of-plane ($c$) direction. The emitted THz pulses (also $c$-polarized) were collimated with a parabolic mirror and refocused on a 1-mm-thick ZnTe crystal to perform electro-optic sampling, directly yielding THz electric field traces in time domain.

In Figs. 1e,1f we report selected time traces of the emitted THz field, measured in both YBCO samples investigated in this study (see Supplemental Material [41] for extended data sets and Ref. [42] for source data). The temperatures at which these data were taken are displayed as full circles in the phase diagrams in Figs. 1b,1c. For completeness, we also show in Fig. 1d the emission traces taken in LBCO 9.5% and already reported in Ref. [26], with the corresponding phase diagram in Fig. 1a. The vertical scales in Fig. 1d-1f have been kept the same, allowing a quantitative comparison between the emission amplitudes in different compounds.

As expected, we observed effectively no response in YBCO 6.48, a compound that shows a weak CDW phase [39] and for which we measured only a temperature-independent single cycle emission just above the noise level (Fig. 1e). This result is very similar to what was reported for optimally-doped $La_{1.84}Sr_{0.16}CuO_4$ (no stripe order) and for $La_{1.885}Ba_{0.115}CuO_4$ (quasi-static stripes) in Ref. [26]. This weak single cycle emission, not relevant for the discussion in this paper, is tentatively attributed to the Dember field generated at the sample surface by photocarrier electron-hole separation [43].

In contrast, at the doping levels corresponding to the most robust CDW (YBCO 6.6, see Fig. 1f), we found a significantly stronger response. Here, at the lowest temperature ($T \ll T_C$), the emitted field is already a factor $\sim$5 larger than in YBCO 6.48. Unlike in LBCO 9.5%, in which the signal decreased for temperatures approaching $T_C$ from below, following the amplitude of the stripes in that compound, in YBCO 6.6 the response grows dramatically with increasing temperature, reaching its maximum near $T_C$, and then reducing abruptly for $T > T_C$. The time traces are also qualitatively different from the coherent multi-cycle emission observed in LBCO 9.5% (Fig. 1d), showing instead a couple of cycles at most in YBCO 6.6 (Fig. 1f).

For a more detailed comparison between the emission properties of these two cuprates, we report in Fig. 2 Fourier transforms of the time traces of Fig. 1d,1f [41]. Beyond the systematically narrower bandwidth found in LBCO 9.5% [41], the temperature dependencies are very different. On the one hand, in LBCO 9.5% the emission frequency progressively redshifts and its amplitude decreases with increasing temperature, disappearing across the transition. By contrast, in YBCO 6.6 we found (i) a gradual growth of the overall emission amplitude as temperature was increased, with a peak near $T_C$ and (ii) a progressive redshift of the peak frequency, similar to that of LBCO 9.5%, with the appearance of a second frequency component at twice the main emission frequency. Note that the Gaussian fits shown in Fig. 2b for data at $T = 0.95T_C$ and $T \simeq T_C$ were performed by constraining the peak frequencies to be $\omega$ and $2\omega$.

In Figure 3 we report a detailed temperature dependence of the peak emission frequencies for different excitation fluences. As in Ref. [26], we show a comparison with the Josephson Plasma Resonance measured at equilibrium with time-resolved THz spectroscopy in the same samples. Its exact frequency was determined by fitting the experimental reflectivity with a Josephson plasma model [26,41].

In both LBCO 9.5% and YBCO 6.6 the main emission frequency traces the temperature dependence of the Josephson Plasma Resonance, approaching it more and more closely for progressively lower fluences but always remaining at values slightly lower than $\omega_{JPR}$ by about 10%. This result confirms that in YBCO 6.6 the CTE mechanism must involve the optical excitation of Josephson plasmons, as similarly discussed for LBCO [26]. Moreover, the evidence of a systematic redshift of $\omega_{CTE}$ with respect to $\omega_{JPR}$ suggests that THz emission in both materials presumably originates from surface modes, whose dispersion curve lies entirely below the bulk Josephson Plasma Resonance (see Ref. [26] for further details on the proposed mechanism).

Even harmonics in the emission spectrum of YBCO 6.6 could originate from a large-amplitude excitation of anharmonic plasmons. This effect should be prohibited in a centrosymmetric layered structure, but could potentially be activated in the presence of a charge order-induced inversion symmetry breaking [26]. However, this mechanism driven by plasmon anharmonicities appears unlikely when looking at the amplitude scaling of the CTE. In the Supplemental Material [41] we plot the amplitude of the $2\omega_{JPR}$ peak against that of the peak at $\omega_{JPR}$ for various data sets taken at different temperatures and pump fluences. This shows a linear dependence and does not connect to a nonlinear plasmon emission, for which one would expect quadratic scaling [44].

## III. Model for coherent terahertz emission

We argue here that the anomalous linear scaling of second harmonic emission is a reporter of superconducting fluctuations. To substantiate this argument, we introduce a model that describes CTE from Josephson plasmons in cuprates with coexisting superconductivity and CDW. First, we consider the time derivative of the Josephson current, $E_{CTE}(t) \propto \frac{\partial}{\partial t} j_{JPR}(t)$, as the source of CTE. According to the first Josephson equation, $j_{JPR}(t) = \omega_{JPR}^2 \sin\theta(t)$, which in the limit of small Josephson phase distortions is expressed as $j_{JPR}(t) \simeq \omega_{JPR}^2 \theta(t)$. Thus, the CTE amplitude scales linearly with the Josephson phase velocity, $\dot{\theta}(t)$, as $E_{CTE} \propto \frac{\partial}{\partial t} j_{JPR}(t) \simeq \omega_{JPR}^2\, \dot{\theta}(t)$. Through the second Josephson equation, $\dot{\theta}(t) = \frac{2eV_J(t)}{\hbar}$, where $V_J(t)$ is the voltage drop across the junction, we obtain a direct proportionality between the emitted THz field, $E_{CTE}(t)$, and the electric field inside the Josephson junction, $E_J(t) = \frac{V_J(t)}{d}$ (here $d$ is the junction size).

Based on our previous work [26,31,32], and having simplified in part the model by leaving out the full analysis of surface modes, we assume that the inherent symmetry breaking by incommensurate CDWs in LBCO 9.5% and YBCO 6.6 enables a Raman excitation process of Josephson plasmons. In this process, the rectified force of two optical photons from the drive induces coherent Josephson plasma oscillations with a strength that is directly proportional to the CDW intensity. Following the derivation reported in the Supplemental Material [41], the frequency-dependent CTE amplitude generated through this mechanism can be expressed as:

$$\left|E_{CTE}^{JPR}(\omega,T)\right| = \left|\frac{I_{CDW}(T)\omega^2(\omega^2+i\gamma_{JPR}\omega)}{-\left(\omega^2+i\gamma_{JPR}\omega-\omega_{JPR}^2(T)\right)^2}\right| E_{opt}E_{opt}, \tag{1}$$

where $I_{CDW}(T)$ is the temperature dependent CDW intensity [27,39], $\gamma_{JPR}$ is a JPR damping coefficient, and $E_{opt}$ denotes the field strength of the optical drive.

Notably, this CTE process is always enabled as long as the incommensurate CDW order is present, a condition that is met throughout the superconducting phase for both LBCO and YBCO compounds studied here. However, while YBCO 6.6 (for which $T_{CDW} \gg T_C$) exhibits a robust CDW phase across the superconducting transition [39], $I_{CDW}(T)$ becomes vanishingly small in LBCO 9.5% for $T \lesssim T_C = T_{CDW}$ [27], thus leading to the strong CTE suppression near $T_C$ observed for this compound.

We next consider an additional contribution to CTE which originates from *second-order* Raman processes involving the excitation of *pairs* of Josephson plasmons, *i.e.*, Josephson *bi-plasmons*, which appear as coherent Josephson plasma oscillations at $\omega = 2\omega_{JPR}$[31,45,46,47,48], in a way similar in many aspects to bi-phonon excitations [49]. The strength of this new term depends critically on the amplitude of superconducting fluctuations and is therefore strongly enhanced close to $T_C$. In addition, its amplitude scales quadratically with the drive electric field, *i.e.*, linearly with pump fluence, as observed for the $2\omega_{JPR}$ peak in our experiment on YBCO 6.6 [41].

In the Supplemental Material [41], we report an explicit derivation for this second CTE channel, according to which the emitted THz field amplitude is expressed as:

$$\left|E_{CTE}^{fluct}(\omega,T)\right| = \left|A\,\frac{N_{fluct}(T)\omega_{JPR}^2(T)}{\omega^2+2i\gamma_{JPR}\omega-4\omega_{JPR}^2(T)}\,\frac{I_{CDW}(T)\omega^2(\omega^2+i\gamma_{JPR}\omega)}{\left(\omega^2+i\gamma_{JPR}\omega-\omega_{JPR}^2(T)\right)^2}\right| E_{opt}E_{opt}, \qquad (2)$$

where $N_{fluct}(T)$ denotes the temperature dependent amplitude of the superconducting fluctuations and $A$ is a scaling factor. Note that in this analysis we have ignored the bilayer structure of YBCO [50,51,52,53,54,55,56,57,58] and approximated the Josephson dynamics as that of a single layered cuprate, like LBCO (see Supplemental Material [41] for validation of this approach).

In Figure 4 we show calculated emission spectra for both compounds at selected temperatures, highlighting separately the different components $\left|E_{CTE}^{JPR}(\omega,T)\right|$ (a,d) and $\left|E_{CTE}^{fluct}(\omega,T)\right|$ (b,e), that contribute to the overall radiated spectrum, $|E_{CTE}^{tot}(\omega,T)| = \left|E_{CTE}^{JPR}(\omega,T) + E_{CTE}^{fluct}(\omega,T)\right|$ (c,f). For $T \ll T_C$ in both LBCO 9.5% and YBCO 6.6 the calculated emission is dominated by the mean field (single-plasmon) term, since

$N_{fluct}(T)$ is vanishingly small. Close to $T_C$ instead, the YBCO 6.6 spectrum is almost entirely attributable to the fluctuating (bi-plasmon) component and shows a prominent peak at $2\omega_{JPR}$ accompanied by a sub-harmonic contribution at $\omega_{JPR}$. In LBCO 9.5% on the other hand, superconducting fluctuations, although present at $T \lesssim T_C$, are silent, because $I_{CDW}(T)$ vanishes at $T_C$ [27], causing the material to lack the prerequisite to radiate.

## IV. Comparison with the experimental data and discussion

In Figure 5 we directly compare these predictions to experimental data. We extracted experimental values of $I_{CDW}(T)$ for both compounds from literature [27,39], and estimated $\omega_{JPR}(T)$ and $\gamma_{JPR}(T)$ from our THz spectroscopy measurements at equilibrium [41]. Using $N_{fluct}(T)$ and $A$ as the only fit parameters, which we tuned to the response of YBCO and then kept fixed for LBCO, our model was able to consistently and quantitatively reproduce both the enhancement of peak emission and the appearance of a $2\omega_{JPR}$ contribution near $T_C$ in YBCO 6.6, as well as the absence of these features in LBCO 9.5%.

Our experiment reveals important new aspects of CTE from cuprates with coexisting superconductivity and charge-density-wave ordering. Firstly, we see confirmation that the presence of fluctuating and/or incommensurate charge order is a fundamental ingredient to observe this phenomenon by providing the necessary inversion symmetry breaking. The new data on YBCO 6.6, showing a systematic redshift of the emission peak frequency with respect to the Josephson Plasma Resonance at equilibrium, are compatible with the idea introduced for LBCO 9.5% in Ref. [26] that the emission likely occurs via the excitation of surface Josephson plasmons rather than bulk plasma polaritons. The new observation here concerns the anomalous enhancement of CTE near the superconducting transition as well as the appearance of a large second harmonic peak. This $2\omega_{JPR}$ contribution exhibits an anomalous linear scaling with pump fluence, which is consistent with the excitation of pairs of Josephson plasmons, as reproduced in detail by our model.

We have shown here that CTE probes quantum fluctuations of the superconducting condensate at $T \lesssim T_C$, where the impulsive Raman mechanism that triggers emission is boosted by these fluctuations. Qualitatively, the observations reported here are a nonlinear analogue of critical opalescence, in which linear light scattering is enhanced by critical fluctuations. Given the indications that coherent THz emission in the superconducting state may be mediated by the excitation of surface rather than bulk Josephson plasmons, we envisage complementing far-field spectroscopy with other techniques sensitive to emission in the near field, such as scanning near-field optical microscopy in the terahertz range (THz-SNOM) [33,59,60,61,62]. Such an approach could provide confirmation of the excitation of surface modes as well as shed new light on the role of inhomogeneities in the coherent emission process.

## Data Availability Statement

The data that support the findings of this article are openly available [42].

## Acknowledgments

We acknowledge support from the Max Planck-New York Center for Non-Equilibrium Quantum Phenomena and from the Deutsche Forschungsgemeinschaft (DFG, German Research Foundation) via the Cluster of Excellence CUI: Advanced Imaging of Matter (EXC 2056, project ID 390715994). E. Demler acknowledges support from AFOSR-MURI: Photonic Quantum Matter award FA95501610323, DARPA DRINQS, the ARO grant "Control of Many-Body States Using Strong Coherent Light-Matter Coupling in Terahertz Cavities". Work at Columbia University was entirely supported by the Center on Precision-Assembled Quantum Materials, funded through the US National Science Foundation (NSF) Materials Research Science and Engineering Centers (award no. DMR-2011738).

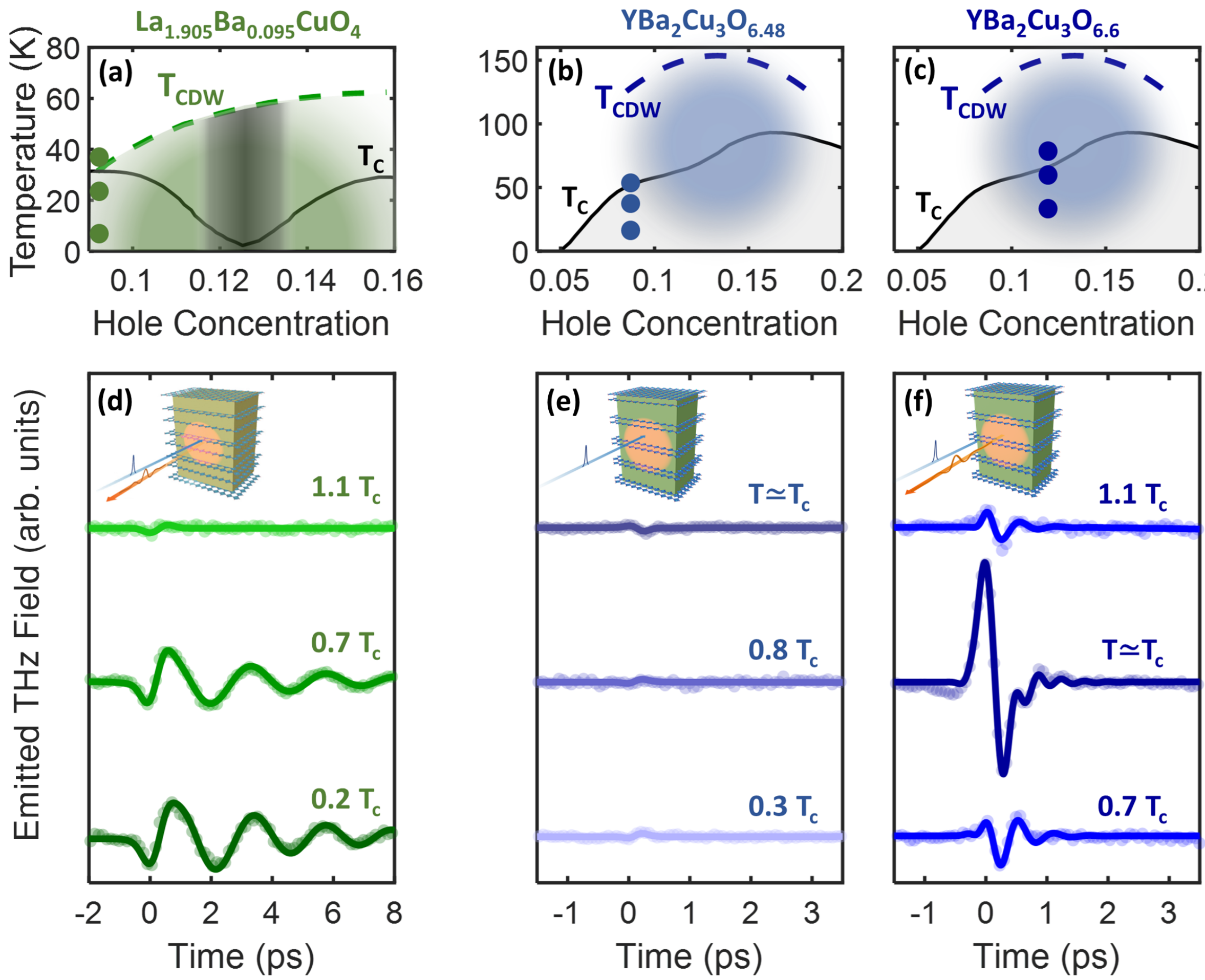

**Figure 1. (a-c)** Temperature-doping phase diagrams of the compounds reported in the present study. $T_{CDW}$ and $T_C$ stand for the charge-density-wave ordering and the superconducting critical temperature, respectively [27,39]. **(d-f)** Time-dependent THz emission traces measured at the temperatures indicated by full circles in (a-c). Solid lines represent multi-component fits [41] while experimental data are displayed as circles. The vertical scales in the three panels are mutually calibrated. Through a calibration with a ZnTe crystal mounted on the same sample holder, we estimate the peak THz fields in (d) and (f) to be no larger than ~10 V/cm. Data in (d) have been taken with a pump fluence of ~2.5 mJ/cm² [26], while those in (e-f) with ~5 mJ/cm². Insets: Experimental geometry. Near-infrared (NIR) pump pulses, with typical duration of ~100 fs, are shone at normal incidence onto an *ac*-oriented sample surface, with polarization parallel to the *c* axis (*i.e.*, perpendicular to the Cu-O planes). As a result of photoexcitation, *c*-polarized THz radiation is emitted.

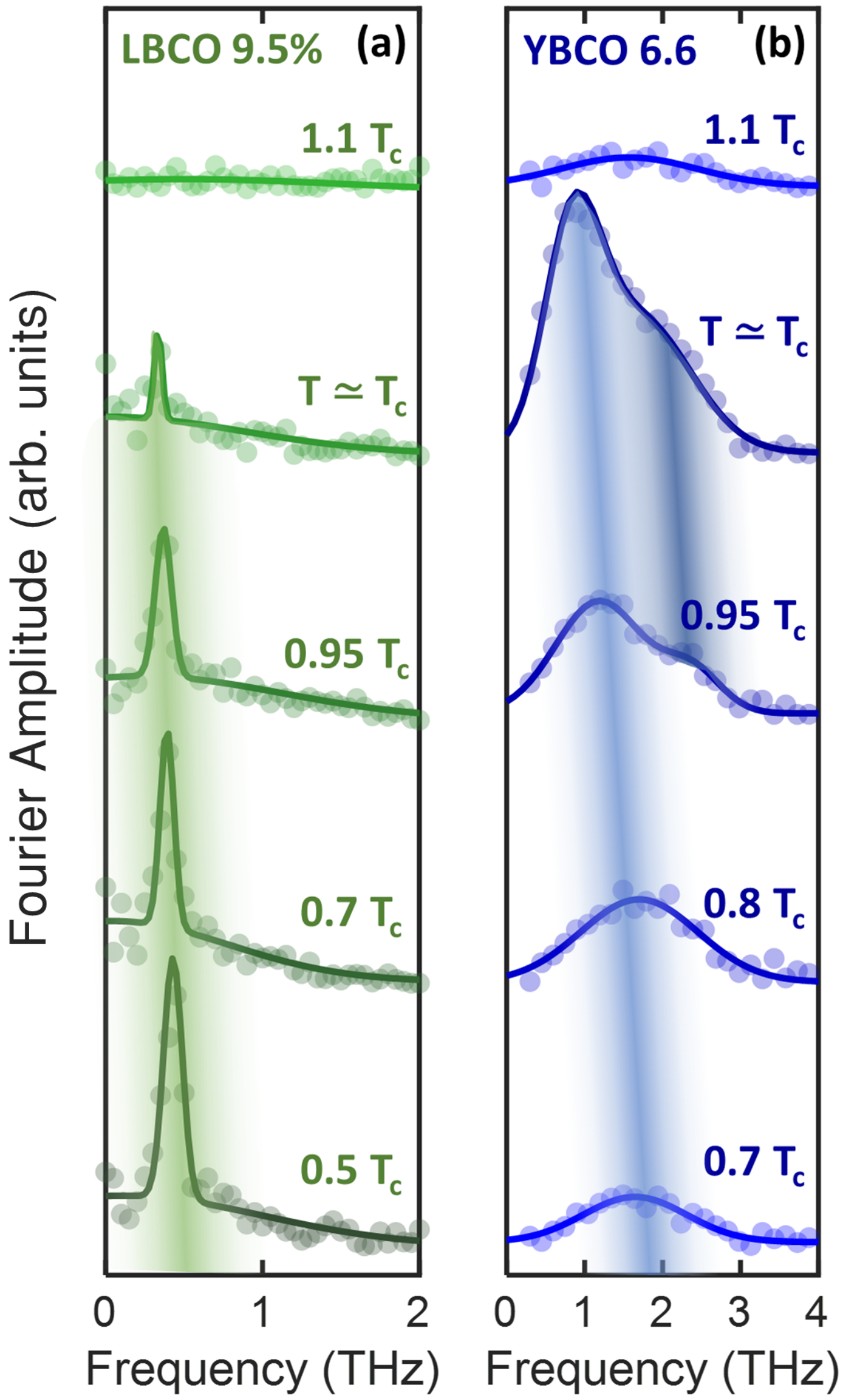

**Figure 2.** Fourier transforms (circles) of the time-domain traces of Fig. 1, shown for both compounds for which we detected a sizeable THz emission, namely $La_{1.905}Ba_{0.095}CuO_4$ **(a)** and $YBa_2Cu_3O_{6.6}$ **(b)**. Solid lines are multi-Gaussian fits while full circles indicate the experimental data. Spectra are reported at selected temperatures above and below the superconducting $T_C$ ($T_C$ = 33 K for $La_{1.905}Ba_{0.095}CuO_4$ and $T_C$ = 63 K for $YBa_2Cu_3O_{6.6}$). Data in (a) were taken with a pump fluence of ~2.5 mJ/cm$^2$ [26], while those in (b) with ~5 mJ/cm$^2$. The shadings track the temperature dependence of the peak emission frequency. Only for $YBa_2Cu_3O_{6.6}$ a second contribution at twice the fundamental frequency appears close to $T_C$.

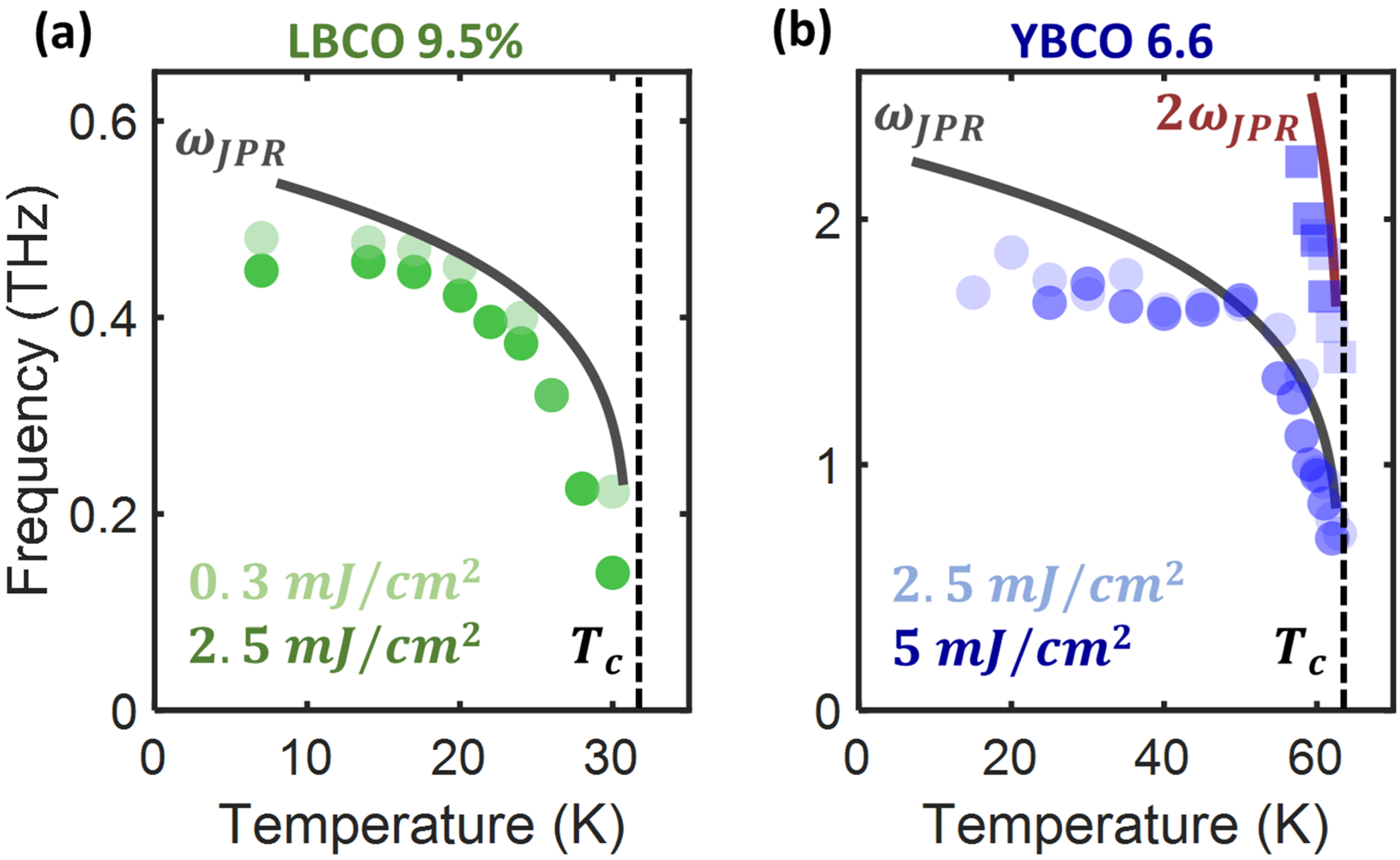

**Figure 3.** Comparison between equilibrium Josephson Plasma Resonance (full black lines) and emission peak frequencies (circles) measured for different excitation fluences (see legend) in $La_{1.905}Ba_{0.095}CuO_4$ [26] and $YBa_2Cu_3O_{6.6}$. $\omega_{JPR}(T)$ was determined by shining a weak, $c$-polarized broadband THz pulse at normal incidence onto the sample surface and detecting the electric field profile of the same THz pulse after reflection. The data were then fitted with a Josephson plasma model [41]. In panel (b) we also compare $2\omega_{JPR}(T)$ (red line) with the second harmonic emission peak frequencies (blue squares), detected in $YBa_2Cu_3O_{6.6}$ at $T \lesssim T_C$.

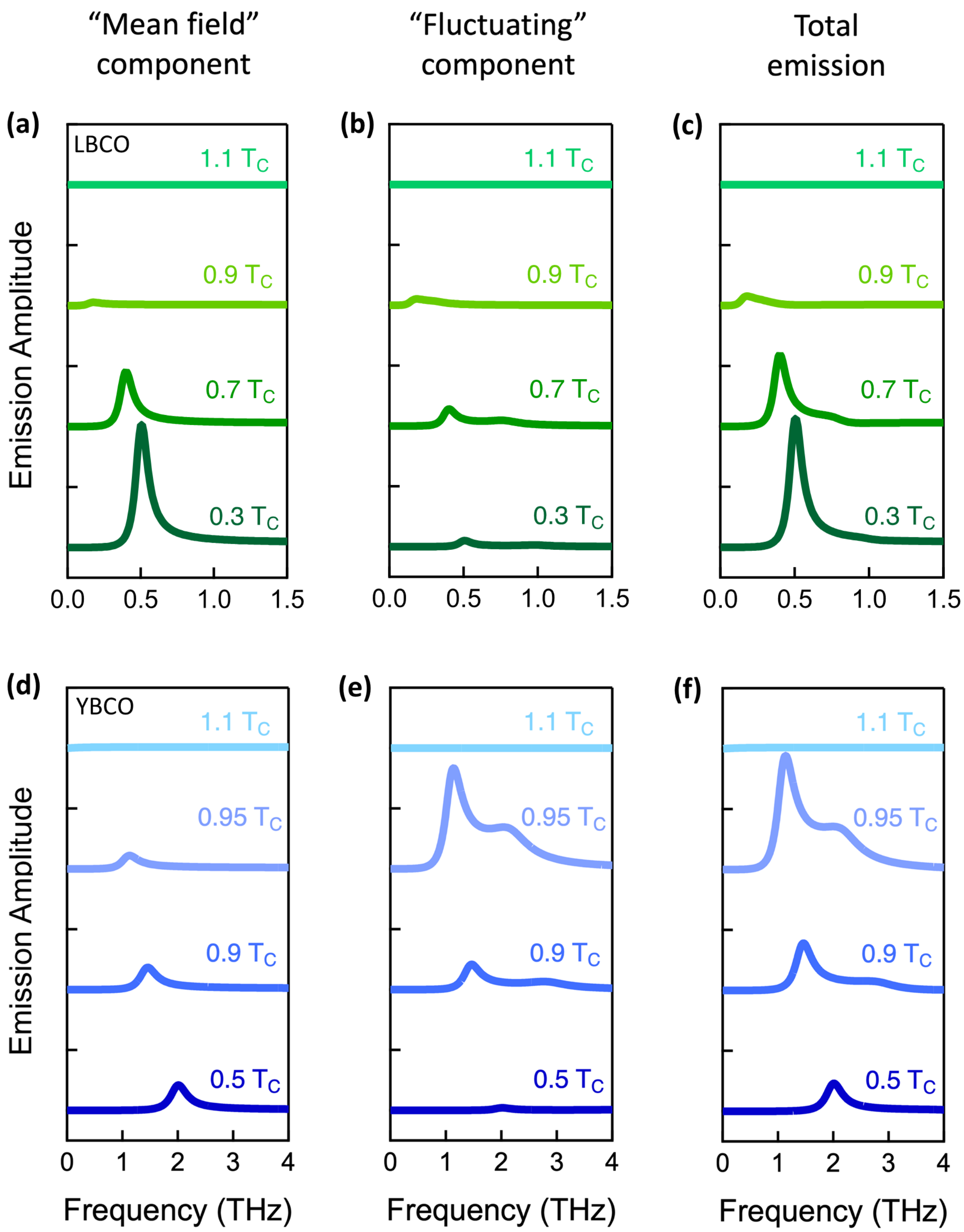

**Figure 4.** THz emission spectra from LBCO 9.5% **(a-c)** and YBCO 6.6 **(d-f)** calculated with the model described in the main text at selected temperatures below and above the superconducting $T_C$. The different panels show the mean field (single-plasmon) contribution to the emitted field (a,d), the bi-plasmon term related to superconducting fluctuations (b,e), and the sum of the two, which gives rise to the total emission (c,f).

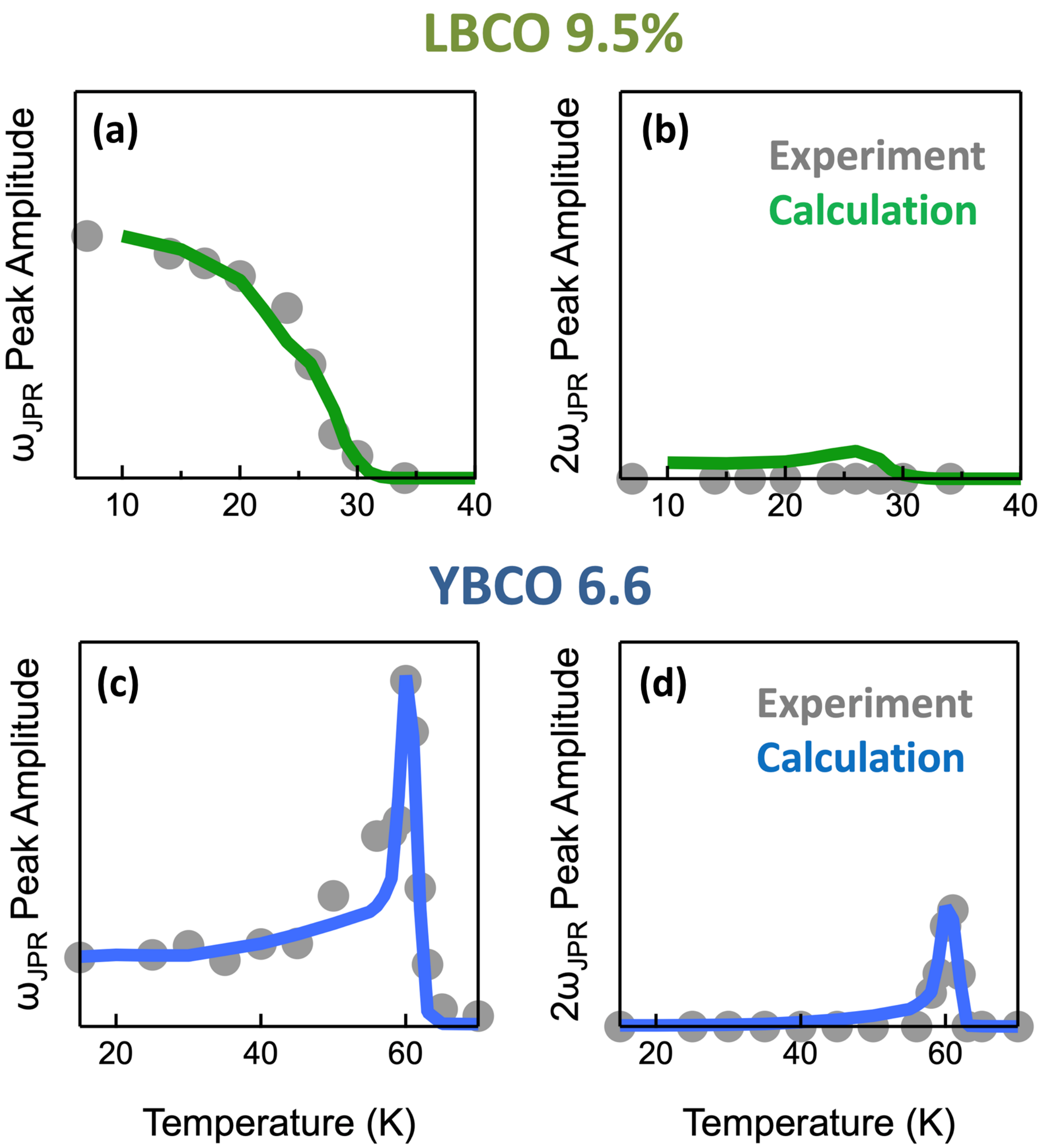

**Figure 5.** Comparison between the temperature dependent peak emission amplitude at $\omega_{JPR}$ **(a,c)** and $2\omega_{JPR}$ **(b,d)** in both LBCO 9.5% and YBCO 6.6, measured experimentally (gray circles) and predicted by the model (full lines). The experimental data in (a-b) were taken with a pump fluence of ~2.5 mJ/cm$^2$, while those in (c-d) with ~5 mJ/cm$^2$.

## REFERENCES (Main Text)

[68] J. Yuan, L. Shi, L. Yue, B. Li, Z. Wang, S. Xu, T. Xu, Y. Wang, Z. Gan, F. Chen, Z. Lin, X. Wang, K. Jin, X. Wang, J. Luo, S. Zhang, Q. Wu, Q. Liu, T. Hu, R. Li, X. Zhou, D. Wu, T. Dong, and N. L. Wang, *Dynamical interplay between superconductivity and pseudogap in cuprates as revealed by terahertz third-harmonic generation spectroscopy*, Sci. Adv. **10**, eadg9211 (2024).

# Enhanced coherent terahertz emission from critical superconducting fluctuations in $YBa_2Cu_3O_{6.6}$

D. Nicoletti[1], M. Rosenberg[1], M. Buzzi[1], M. Fechner[1], Y. Liu[2], S. Nakata[2], B. Keimer[2], R. A. Vitalone[3], D. N. Basov[3], P. E. Dolgirev[4], E. Demler[5], M. H. Michael[1], and A. Cavalleri[1,6]

[1] *Max Planck Institute for the Structure and Dynamics of Matter, 22761 Hamburg, Germany*

[2] *Max Planck Institute for Solid State Research, 70569 Stuttgart, Germany*

[3] *Department of Physics, Columbia University, New York, NY 10027, USA*

[4] *Department of Physics, Harvard University, Cambridge, Massachusetts 02138, USA*

[5] *Institute for Theoretical Physics, ETH Zurich, 8093 Zurich, Switzerland*

[6] *Department of Physics, Clarendon Laboratory, University of Oxford, Oxford OX1 3PU, United Kingdom*

# Supplemental Material

## S1. Additional data sets

In this Section, we report extended data sets, which complement the time traces and spectra in Fig. 1-2 of the main text. These were acquired on $YBa_2Cu_3O_{6.6}$ (YBCO 6.6) for three different excitation fluences, namely 2.5, 5, and 10 mJ/cm$^2$, and as a function of temperature. From these extended data sets, the experimental points shown in Fig. 3, 5 of the main text and Fig. S5, S6 were derived.

In Fig. S1 we show time traces of emitted fields measured at 5 different temperatures for each fluence, while in Fig. S2 we report the corresponding spectra extracted by Fourier transform.

The salient features of the emission from YBCO 6.6 as a function of temperature, *i.e.*, the enhancement of the emitted THz field near the superconducting $T_C$, as well as the appearance of a second harmonic peak in the spectrum, are present in all data, regardless of the applied fluence.

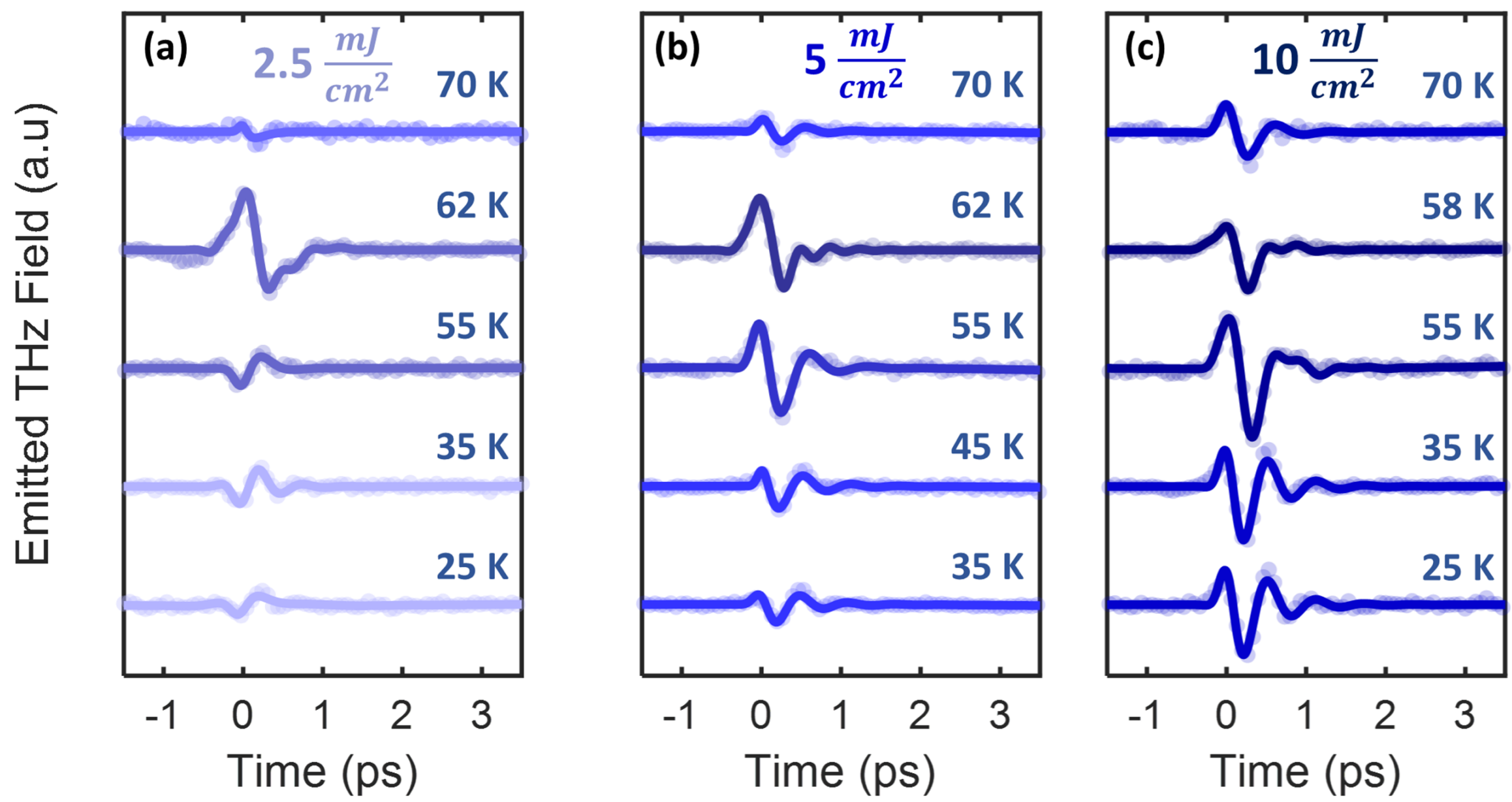

**Figure S1.** Extended sets of THz emission traces in the time domain measured in $YBa_2Cu_3O_{6.6}$ ($T_C$ = 63 K) as a function of temperature and for different excitation fluences of 2.5 mJ/cm$^2$ **(a)**, 5 mJ/cm$^2$ **(b)**, and 10 mJ/cm$^2$ **(c)**. Solid lines represent multi-component fits (see Section S2) while the experimental data are displayed as circles. The vertical scales in (b,c) are mutually calibrated, while data in (a) have been multiplied by a factor of 2 for better visualization.

Differences are found instead in the temperature range corresponding to the maximum emission amplitude, which appears to progressively reduce by a few degrees with increasing fluence. This effect is presumably due to a pump-induced heating, which acquires a more important role as the energy deposited on the sample increases.

Note that in most of the analysis reported in the main text we chose to use the 5 mJ/cm$^2$ data, as those showed both a clear enhancement of the emission near $T_C$ and a strong second harmonic peak over a sufficiently wide temperature range.

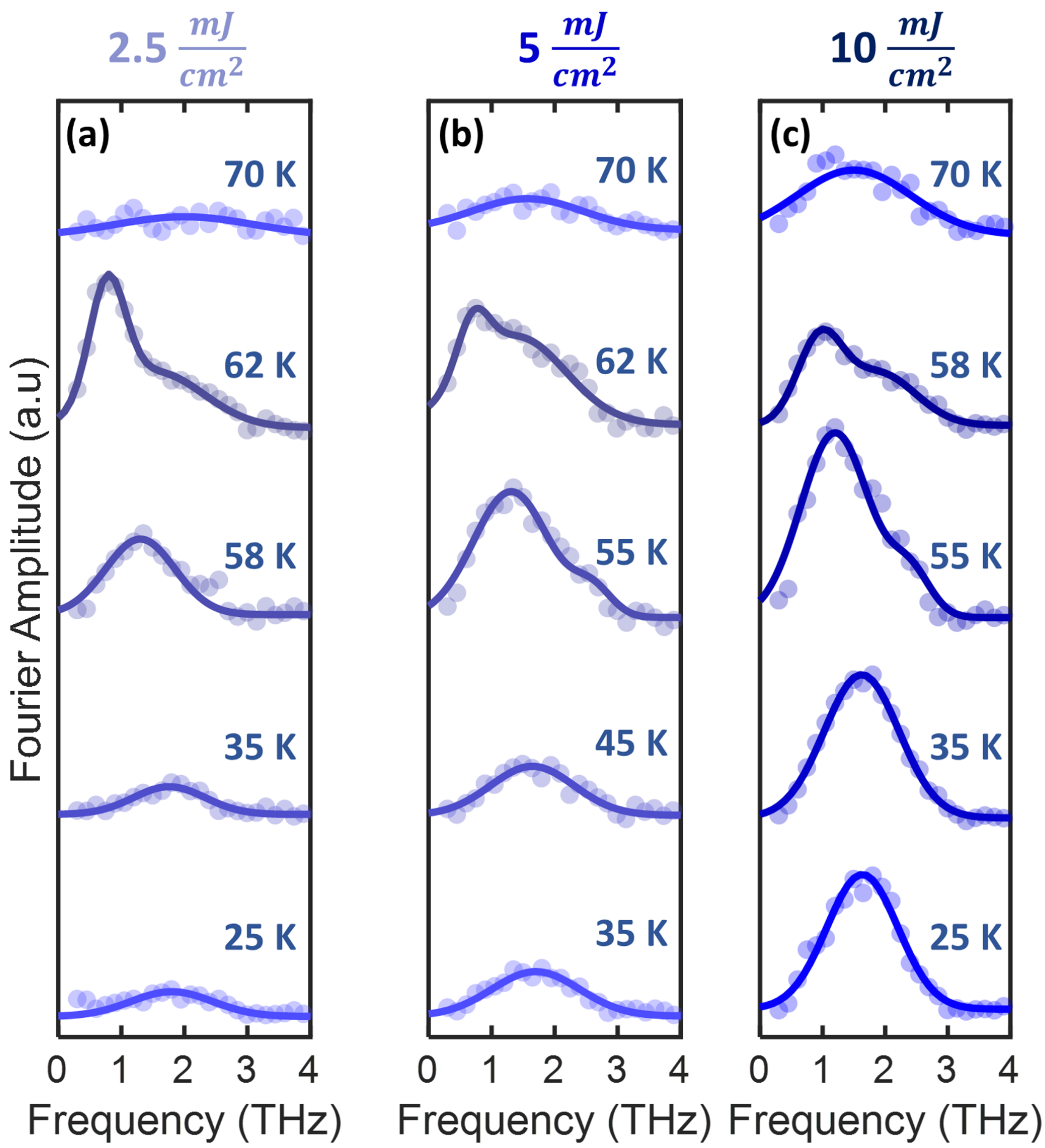

**Figure S2.** Fourier transforms of the time domain traces of Fig. S1, measured in $YBa_2Cu_3O_{6.6}$ ($T_C$ = 63 K) as a function of temperature and for different excitation fluences of 2.5 mJ/cm$^2$ **(a)**, 5 mJ/cm$^2$ **(b)**, and 10 mJ/cm$^2$ **(c)**. Solid lines are multi-Gaussian fits while full circles indicate the experimental data. The vertical scales in (b,c) are mutually calibrated, while data in (a) have been multiplied by a factor of 2 for better visualization.

## S2. Fitting model for the emission traces in time domain

All time-dependent experimental curves shown in Fig. 1 of the main text and Fig. S1 of the Supplemental Material, were fitted using the formula:

$$E_{THz}(t) = A_0 e^{-(t/2\tau_0)^2} \cos(\omega_0 t + \varphi_0) + A_1[1 + \mathrm{erf}(t/\tau_1)]e^{-\gamma_1 t} \cos[\omega_{CTE} t + \varphi_1] +$$
$$+ A_2[1 + \mathrm{erf}(t/\tau_2)]e^{-\gamma_2 t} \cos[2\omega_{CTE} t + \varphi_2]$$

Here, $A_0$, $\tau_0$, $\omega_0$, and $\varphi_0$ are the amplitude, Gaussian width, central frequency, and phase of a weak, single-cycle component around time zero, present in all materials and for all temperatures below and above $T_C$, that we tentatively assign to the Dember field generated at the sample surface by photo-carrier electron-hole separation (see also main text).

$A_1$, $\tau_1$, $\gamma_1$, $\omega_1$, and $\varphi_1$ are amplitude, rise time, decay rate, frequency, and phase of the main THz emission component at $\omega_{CTE} \simeq \omega_{JPR}$, found only in the superconducting state of LBCO 9.5% and YBCO 6.6.

Finally, $A_2$, $\tau_2$, $\gamma_2$, and $\varphi_2$ are amplitude, rise time, decay rate, and phase of the second harmonic emission component at $2\omega_{CTE} \simeq 2\omega_{JPR}$, which we observed in YBCO 6.6 for temperatures near the superconducting $T_C$.

## S3. Equilibrium Josephson plasma frequency of $YBa_2Cu_3O_{6.6}$

In Figure S3 we show the temperature dependence of the equilibrium Josephson Plasma Resonance of YBCO 6.6.

Following the same procedure applied for LBCO in Ref. [i], this was determined by measuring the *c*-axis reflectivity at various temperatures in the superconducting state with THz time-domain spectroscopy and normalizing it by the same quantity measured at $T \gtrsim T_C$. These reflectivity ratios (Fig. S3(a)) were then fitted with a Josephson Plasma model[ii], thus extracting $\omega_{JPR}(T)$ (see Fig. S3(b)), which we found to be in agreement with previous determinations in the literature[iii,iv].

Note that these same data, complemented by $\omega_{JPR}(T)$ of LBCO 9.5% from Ref. [i], have been used in Fig. 3 of the main text (black lines) for a comparison of the equilibrium Josephson Plasma frequencies with the peak emission frequencies of both compounds.

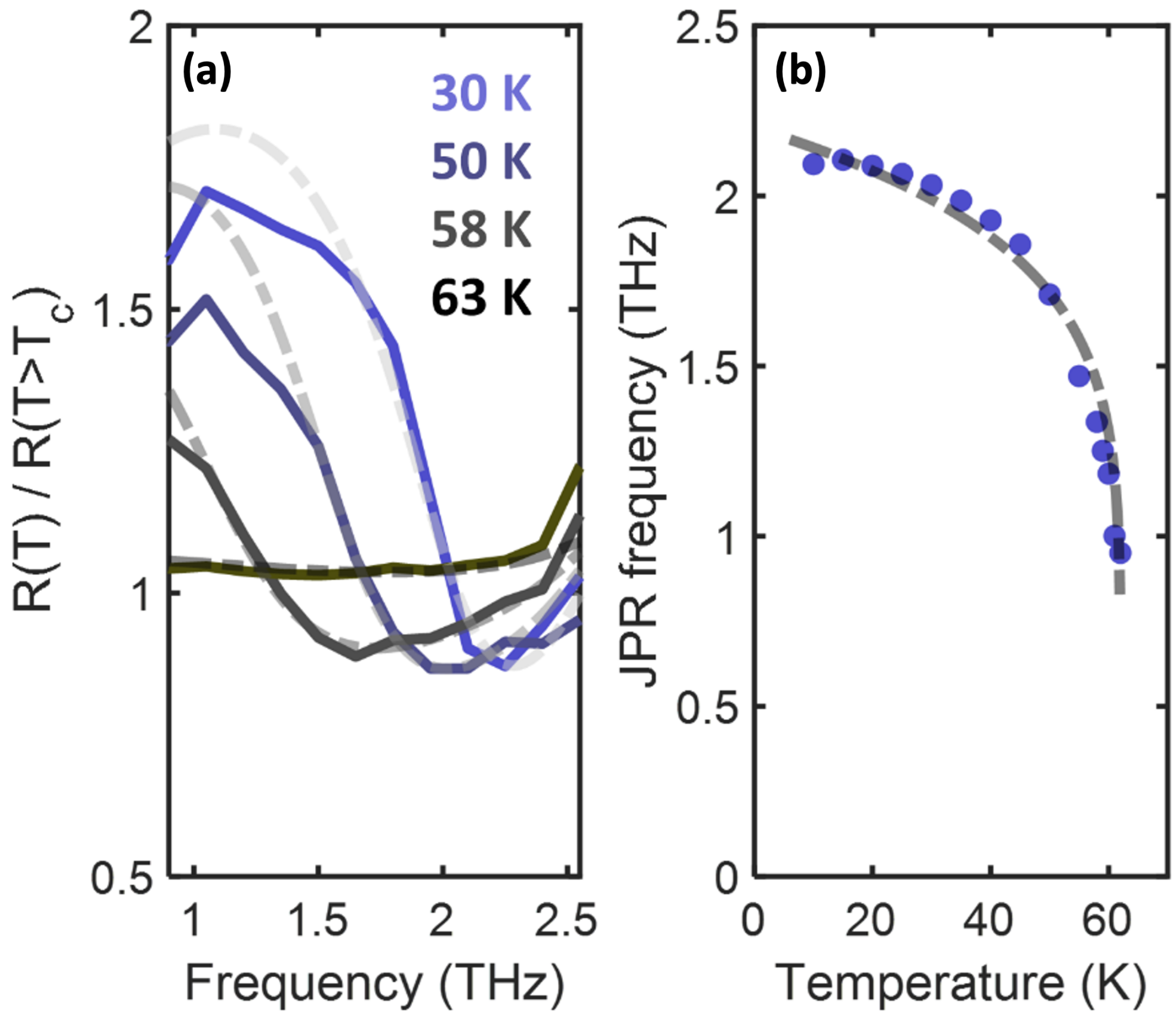

**Figure S3. (a)** Reflectivity of $YBa_2Cu_3O_{6.6}$, normalized by the same quantity measured in the normal state above $T_C$. All spectra were measured by THz time-domain spectroscopy and are reported at selected temperatures below and across the superconducting transition. The reference temperature for normalization was 70 K. The Josephson Plasma Resonance appears here as a sharp edge located at $\omega = \omega_{JPR}$. Dashed grey lines are fits to the data with a Josephson plasma model. **(b)** Temperature dependence of $\omega_{JPR}$ extracted from the fits on the reflectivity ratios in (a). The dashed line is a guide to the eye.

## S4. Analysis of the emission bandwidth

In Fig. S4 we report an analysis of the THz emission bandwidth in various compounds studied in the present work and in Ref. [i]. We compare emission spectra taken at $T \ll T_C$ in LBCO 9.5%, LBCO 15.5%, and YBCO 6.6 (Fig. S4(d-f)), for which only one peak was found in the spectrum, with the energy loss function extracted from the reflectivity measured in the same compounds at equilibrium (see, *e.g.*, Fig. S3), at the same temperature (Fig. S4(a-c)).

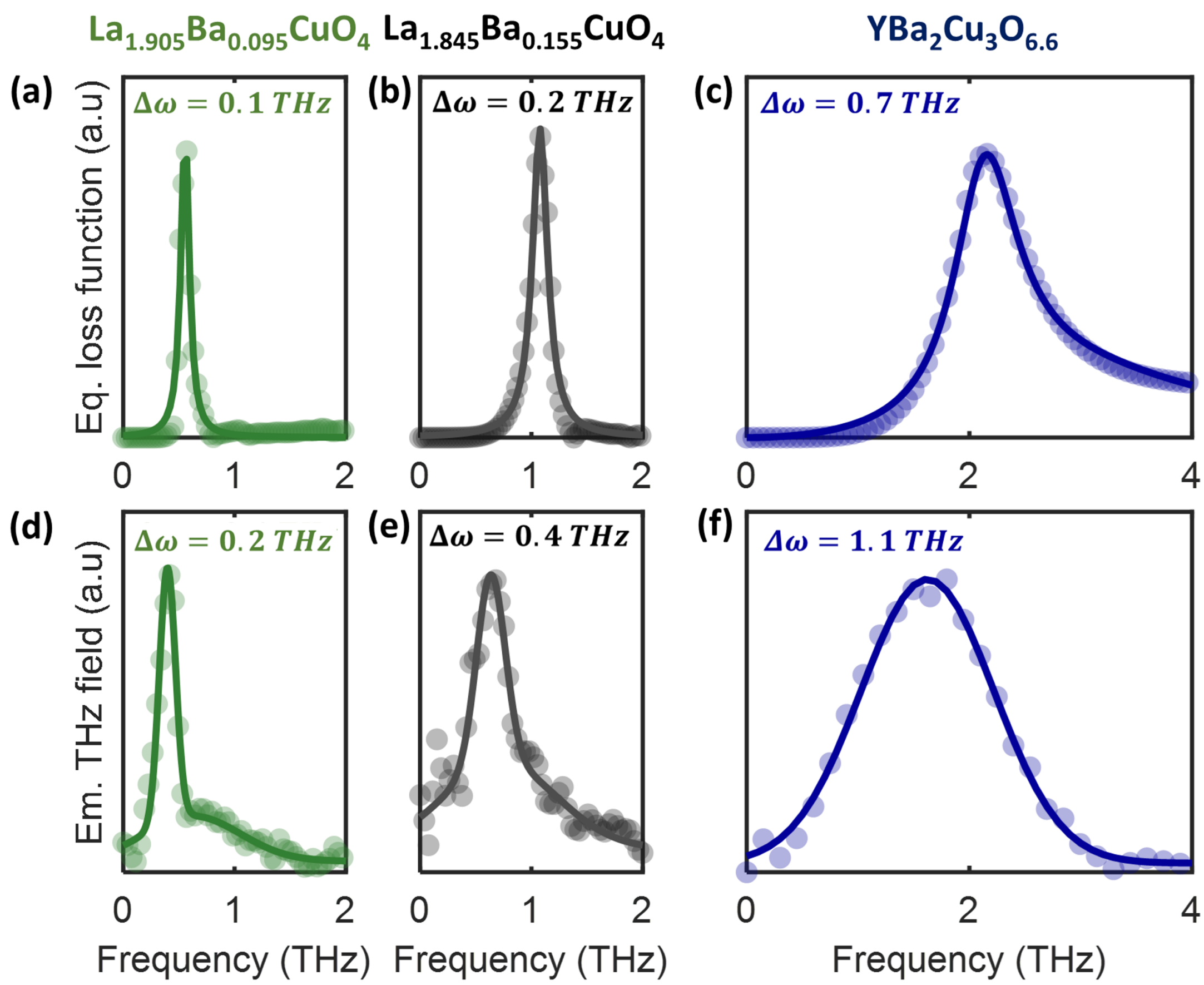

**Figure S4. (a-c)** Energy loss function extracted from the reflectivity ratios reported in Ref. [i] for both LBCO compounds and in Fig. S3 of this manuscript for YBCO 6.6. Circles are data points derived via Kramers-Kronig transformations, while full lines are fits with a Josephson Plasma model[ii]. **(d-f)** Emission spectra measured for the same compounds. Here, circles are experimental data while full lines are Gaussian fits. Data in (d,e) were taken with 2.5 mJ/cm$^2$ fluence, those in (f) with 5 mJ/cm$^2$. The extracted bandwidths are explicitly indicated in all panels. The data in (a,b,d,e) were measured at $T = 10$ K, while those (c,f) at $T = 25$ K, all temperatures well below the superconducting $T_C$, for which only one component in the emission spectrum was found.

Such a comparison allows us to systematically study the connection between the THz emission bandwidth and the intrinsic width of the Josephson Plasma Resonance at equilibrium, which manifests as a peak in the energy loss function.

As seen in Fig. S4, the progressive broadening of the emission spectrum from LBCO 9.5% to YBCO 6.6 via LBCO 15.5% seems to originate from an increase in the width of the Josephson plasmon, which nevertheless remains systematically narrower by a factor of ~2 than the emission bandwidth.

We stress here that no such trend was found instead by a comparison with the charge density wave properties in the various compounds, which in fact have effectively identical correlation lengths, around 50 Å [v,vi].

Based on these observations we conclude that the emission spectrum in charge-ordered cuprates at $T \ll T_C$ appears to be entirely dominated, both in its peak frequency and bandwidth, by the Josephson Plasma Resonance.

## S5. Amplitude scaling of the emission peaks

In Figure S5 we show the amplitude of the $\omega_{JPR}$ emission peak as a function of excitation fluence, measured in $YBa_2Cu_3O_{6.6}$ at $T = 35$ K $\ll T_C$, a temperature for which no second harmonic emission was detected. This shows a linear scaling (i.e., $\propto E_{drive}^2$), consistent with the prediction of the model discussed in Section S6.

In Figure S6 we report the amplitude of the $2\omega_{JPR}$ emission peak as a function of that of the corresponding peak at $\omega_{JPR}$, both extracted by multi-Gaussian fits to the emission spectra measured in $YBa_2Cu_3O_{6.6}$. The plot shows together data acquired at different temperatures near the superconducting transition, and at different excitation fluences (see Section S1 for full data sets). A linear dependence over more than one octave was found, thus indicating that the generation of the second harmonic peak cannot be attributed to a plasmon anharmonicity, for which one would expect instead a quadratic scaling[vii].

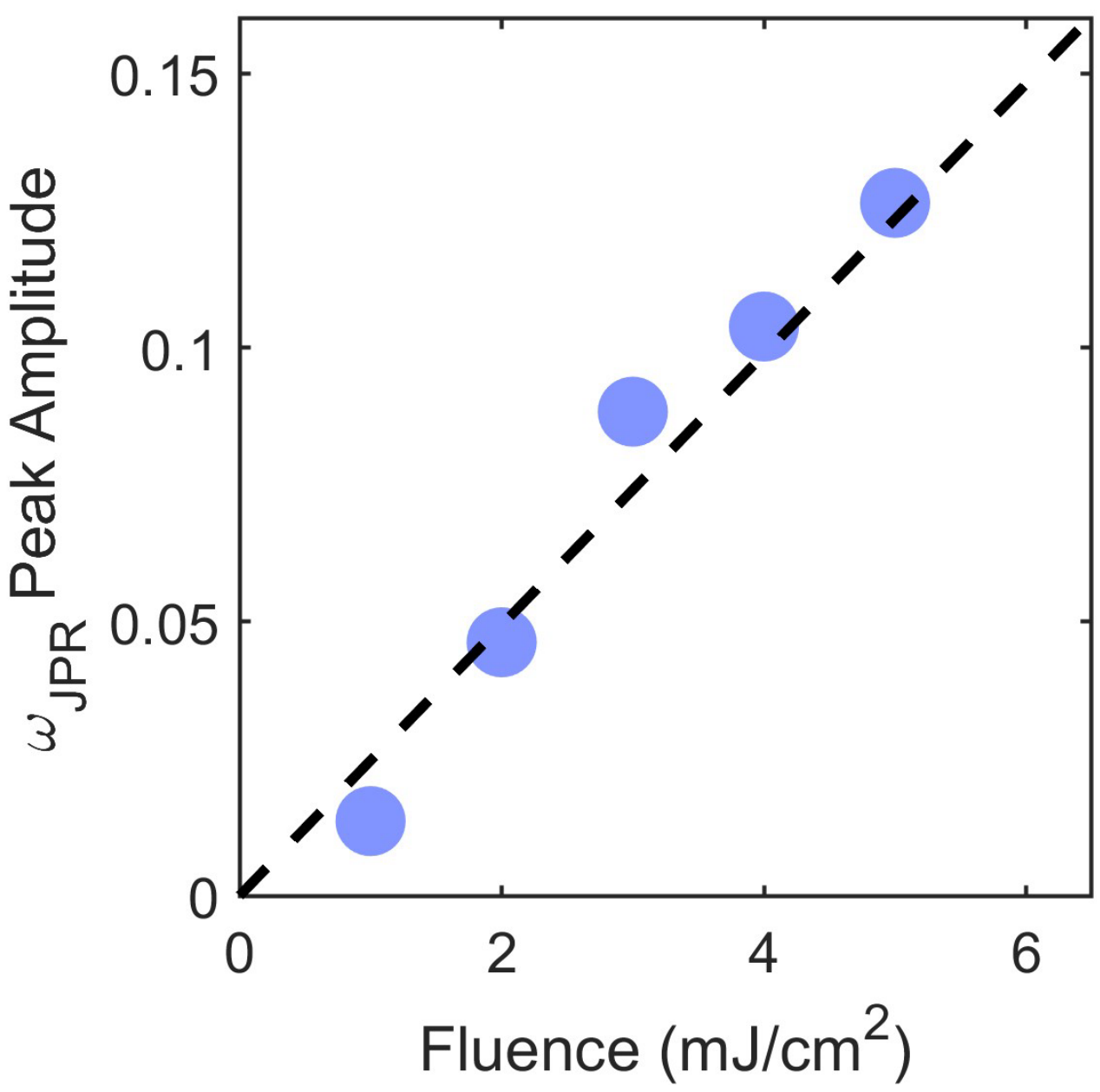

**Figure S5.** Amplitude of the $\omega_{JPR}$ emission peak, extracted by Gaussian fits to the emission spectra measured in $YBa_2Cu_3O_{6.6}$ at $T = 35\,K \ll T_C$, as a function of excitation fluence. The dashed line is a linear fit to the data.

## S6. Model for coherent THz emission activated by superconducting fluctuations

Here, we introduce a model aimed at explaining the observed THz emission close to the Josephson Plasma frequency, $\omega_{JPR}$, as well as a second harmonic contribution at $\sim 2\omega_{JPR}$ in cuprates with coexisting superconductivity and CDW order. As in the theory for photoexcited $La_{2-x}Ba_xCuO_4$ introduced in Refs. [viii,ix], also in the present case the optical drive plays the role of exciting carriers at the cuprate surface. The inversion symmetry breaking responsible for THz emission is originated from the CDW, which gives rise to both zero-momentum and Umklapp shift currents[viii,ix].

Here, we simplify in part the approach of Ref. [viii,ix] by leaving out the full analysis of surface Josephson plasmons, but we take one step forward and introduce a minimal set of assumptions required to explain the appearance of a response at $2\omega_{JPR}$ for $T \lesssim T_C$, which exhibits an anomalous *linear* scaling with the $\omega_{JPR}$ peak amplitude.

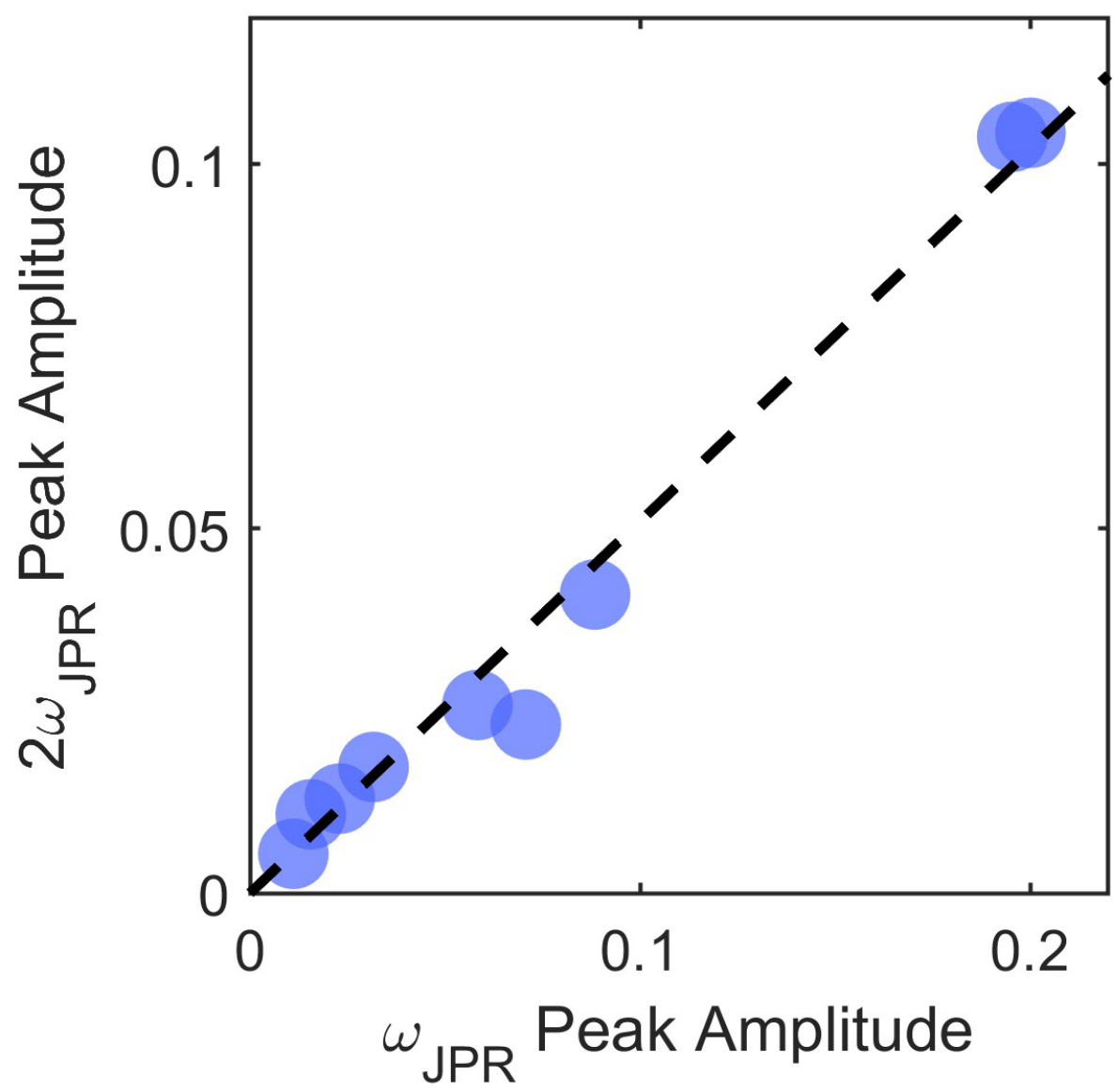

**Figure S6.** Amplitude of the $2\omega_{JPR}$ emission peak as a function of that of the corresponding peak at $\omega_{JPR}$, both extracted by multi-Gaussian fits to the emission spectra measured in $YBa_2Cu_3O_{6.6}$. The plot shows together data acquired at different temperatures near the superconducting transition ($T \simeq 0.97 - 0.98T_C$), and at different excitation fluences. The dashed line is a linear fit to the data.

## S6.1 Josephson plasmons without CDW

The Josephson plasmon model introduced in Refs. [viii,ix] can be described through the Hamiltonian:

$$H_0 = \sum_i \left( \frac{2e^2 d}{\varepsilon_0} \int \mathrm{d}^2 r \pi_i^2(r) - \frac{\Lambda_0 \varepsilon_0}{4e^2 d} \int \mathrm{d}^2 r \cos\theta_i(r) + \frac{\varepsilon_0}{8e^2 d} \sum_j \int \mathrm{d}^2 r L_{ij} \nabla\theta_i(r) \nabla\theta_j(r) \right) \quad (1)$$

Here, $e$ is the electron charge, $\varepsilon_0$ the vacuum permittivity, $d$ the inter-layer separation, and $\Lambda_0 = \omega_{JPR}^2$ the "zero-momentum" superfluid density. $i$ ($j$) and $r$ are indices for the $i$-th ($j$-th) Cu-O layer and a vector coordinate parallel to the Cu-O planes, respectively. $L_{ij}$ is a matrix defining the capacitive coupling between two neighbouring layers, as discussed in Ref. [x]. Finally, $\pi_i(r,t)$ is the conjugate momentum of the gauge invariant Josephson phase $\theta_i$, such that $[\theta_i(r,t), \pi_j(r',t)] = i\delta_{i,j}\delta(r-r')$. The relation between $\pi_i(r,t)$ and the electric field $E_i(r,t)$ is given by:

$$\partial_t\theta_i(r,t) = 2edE_i(r,t) = \frac{4e^2d}{\varepsilon_0}\pi_i(r,t) \Rightarrow \pi_i(r,t) = \frac{\varepsilon_0}{2e}E_i(r,t). \tag{2}$$

We can then rewrite the Hamiltonian (1) in terms of $E_i$ and $\theta_i$ as:

$$H_0 = \sum_i\left(\frac{d\varepsilon_0}{2}\int \mathrm{d}^2r E_i^2(r) - \frac{\Lambda_0\varepsilon_0}{4e^2d}\int \mathrm{d}^2r\cos\theta_i(r) + \frac{\varepsilon_0}{8e^2d}\sum_j\int \mathrm{d}^2r L_{ij}\nabla\theta_i(r)\nabla\theta_j(r)\right) \tag{3}$$

with commutation relations: $[\theta_i(r,t), E_j(r',t)] = i\frac{2e}{\varepsilon_0}\delta_{i,j}\delta(r-r')$. This leads to the equations of motion:

$$(\partial_t^2 + \gamma\partial_t)E_n - \nabla^2 L_{nm}E_m + \Lambda_0 E_n = 0. \tag{4}$$

For $q = 0$, this expression becomes:

$$(\partial_t^2 + \gamma\partial_t)E_n + \Lambda_0 E_n = 0. \tag{5}$$

## S6.2 Josephson plasmons in the presence of a CDW

At this point, we introduce a modification in the Josephson plasmon Hamiltonian to account for the presence of the charge density wave. Equation (3) is rewritten then as:

$$H_{CDW} = \sum_i\left(\frac{d\varepsilon_0}{2}\int \mathrm{d}^2r E_i^2(r) - \frac{\Lambda_i\varepsilon_0}{4e^2d}\int \mathrm{d}^2r\cos\theta_i(r) + \frac{\varepsilon_0}{8e^2d}\sum_j\int \mathrm{d}^2r L_{ij}\nabla\theta_i(r)\nabla\theta_j(r)\right) +$$

$$\sum_i \mathrm{d}^2r\left(dP_{CDW,i}E_i + \alpha\frac{d\varepsilon_0 E^4}{4}\right). \tag{6}$$

Here, we have assumed that the equilibrium state has local electric fields due to the presence of the CDW. These give rise to local dipoles described by the static polarization $P_{CDW}$, but no net currents. Crucially, the $E^4$ nonlinearity, with the amplitude coefficient $\alpha$, becomes a cubic nonlinearity in the presence of the CDW, and is therefore able to generate a second harmonic resonance. The new equations of motion are given by:

$$\partial_t\theta_i = 2edE_i + 2\alpha edE_i^3 + 2ed\frac{P_{CDW,i}}{\varepsilon_0}. \tag{7}$$

In equilibrium, $\partial_t\theta_i = 0$, *i.e.*, no current is flowing. In this condition, the electric field is expressed, for small $\alpha$, as:

$$E_i + \alpha E_i^3 = -\frac{P_{CDW,i}}{\varepsilon_0} \Rightarrow E_i \approx -\frac{P_{CDW,i}}{\varepsilon_0} \tag{8}$$

By expanding around $E_i \approx -\frac{P_{CDW,i}}{\varepsilon_0} + \tilde{E}_i$, we obtain the following Hamiltonian for the CDW state:

$$H_{CDW} = \sum_i \left(\frac{d\varepsilon_0}{2}\int \mathrm{d}^2r\tilde{E}_i^2(r) - \frac{\Lambda_i\varepsilon_0}{4e^2d}\int \mathrm{d}^2r\cos\theta_i(r) + \frac{\varepsilon_0}{8e^2d}\sum_j\int \mathrm{d}^2rL_{ij}\nabla\theta_i(r)\nabla\theta_j(r)\right) + \sum_i \mathrm{d}^2r\left(-\alpha d\varepsilon_0P_{CDW,i}\tilde{E}_i^3\right). \tag{9}$$

Here, the equations of motion for $q = 0$ are given by:

$$\partial_t\theta_i = 2edE_i - 6\alpha edP_{CDW,i}E_i^2 \tag{10}$$

$$\partial_tE_i = \partial_x^2L_{ij}\theta_j - \frac{\Lambda_i}{2ed}\sin(\theta_i) \tag{11}$$

$$\Rightarrow \partial_t^2E_i - \partial_x^2L_{ij}E_j + \Lambda_i\cos(\theta_i)\,E_i = 3\alpha\Lambda_i\cos(\theta_i)\,P_{CDW,i}E_i^2. \tag{12}$$

Now we can work perturbatively in $E_i$ and extract the second harmonic generation response, given a first order response which is expressed as:

$$(\partial_t^2 + \Lambda_i)E_i = 3\alpha\Lambda_iP_{CDW,i}E_i^2. \tag{13}$$

In the CDW state, the Fourier components of $\Lambda_i$ are $\Lambda_0, \Lambda_Q$, and $\Lambda_{-Q}$ (here $Q$ is the CDW wavevector). Those of $P_{CDW,i}$ are $P_{CDW,q=0} = 0$ and $P_{CDW,Q}, P_{CDW,-Q} \neq 0$. In frequency and momentum space, the second order response, $E_q^{(2)}(\omega)$ can be expressed as a function of the first, $E_q^{(1)}$, as:

$$\begin{pmatrix} -\omega^2+\Lambda_0 & \Lambda_Q & 0 \\ \Lambda_{-Q} & -\omega^2+\Lambda_0 & \Lambda_Q \\ 0 & \Lambda_{-Q} & -\omega^2+\Lambda_0 \end{pmatrix} \cdot \begin{pmatrix} E_{-Q}^{(2)} \\ E_{q=0}^{(2)} \\ E_Q^{(2)} \end{pmatrix} = 3\alpha \begin{pmatrix} \Lambda_0 P_{CDW,-Q} \\ \Lambda_Q P_{CDW,-Q} + \Lambda_{-Q} P_{CDW,Q} \\ \Lambda_0 P_{CDW,Q} \end{pmatrix} \left[\left(E_{q=0}^{(1)}\right)^2\right] \tag{14}$$

For $q = 0$, we obtain:

$$E_{q=0}^{(2)} = 3\alpha \left[\left(E_{q=0}^{(1)}\right)^2\right] \frac{(\Lambda_{-Q} P_{CDW,Q} + \Lambda_Q P_{CDW,-Q})(\omega^2 + i\gamma\omega)}{2\Lambda_Q \Lambda_{-Q} - (\omega^2 + i\gamma\omega - \Lambda_0)^2}. \tag{15}$$

Here, $\gamma$ is a phenomenological damping coefficient. The above equation expresses the so-called "mean field" (single plasmon) contribution to the emitted THz field (see Fig. 4 in the main text), which as expected and as verified experimentally[i] (see Fig. S5) scales quadratically with the electric field of the optical drive, *i.e.*, linearly with its intensity.

### S6.3 Bi-plasmon dynamics

We now turn to the definition of a second contribution to the emitted THz field, namely that attributed to superconducting fluctuations, which is expected to be predominant at $T \lesssim T_C$.

In our theoretical model above, we have considered a perturbation theory in the CDW amplitude, defined by the parameters $\Lambda_Q$ and $P_{CDW,Q}$. Consistent with this, we first compute here the squeezed plasmon dynamics in terms of oscillations of $\langle\theta^2\rangle(t)$ in the absence of CDW order, and then consider the role of the CDW in outcoupling the radiation to give rise to THz emission.

The system Hamiltonian can be written as follows:

$$H_0 = \int \frac{\mathrm{d}^2\mathbf{q}\mathrm{d}k_z}{(2\pi)^3} \left[ d\varepsilon_0 E_{\mathbf{q},k_z} E_{-\mathbf{q},-k_z}(r) + \frac{\Lambda_0 \varepsilon_0 \left(1 - \beta E_{drive}^2(t)\right)}{4e^2 d} \theta_{\boldsymbol{q},k_z} \theta_{-\boldsymbol{q},-k_z} + \mathcal{O}(\theta^4) + \right.$$
$$\left. \frac{\varepsilon_0}{4e^2 d} c^2(k_z) \mathbf{q}^2 \theta_{\boldsymbol{q},k_z} \theta_{-\boldsymbol{q},-k_z} \right] \tag{16}$$

where $\mathbf{q}$ and $k_z$ are in- and out-of-plane momenta, $c^2(k_z) = \frac{c^2}{\frac{\lambda_{ab}^2}{d^2}(2-2\cos(k_z d))+1}$ and $\lambda_{ab}$ is the London penetration depth associated with in-plane currents.

Note that in Eq. (16) we have ignored the bilayer structure of YBCO and approximated the Josephson dynamics as that of a single layered cuprate (like LBCO). This approach is justified because the upper (intra-bilayer) Josephson plasmon in YBCO is located at ~14 THz, *i.e.*, far beyond the detection range of our experiment, while the bi-plasmon fluctuation density of states peaks at $\omega = 2\omega_{JPR} \simeq 3-4$ THz. This is shown in Fig. S7, where we report the calculated "bare" density of states of bi-plasmons along with the "fluctuation" density of states. While the former grows with frequency, the latter is expected to be strongly depleted for $\omega \gg 2\omega_{JPR}$. For more details on complications arising from the bilayer nature of YBCO we point the reader to Refs. [xi,xii,xiii,xiv,xv,xvi,xvii,xviii,xix].

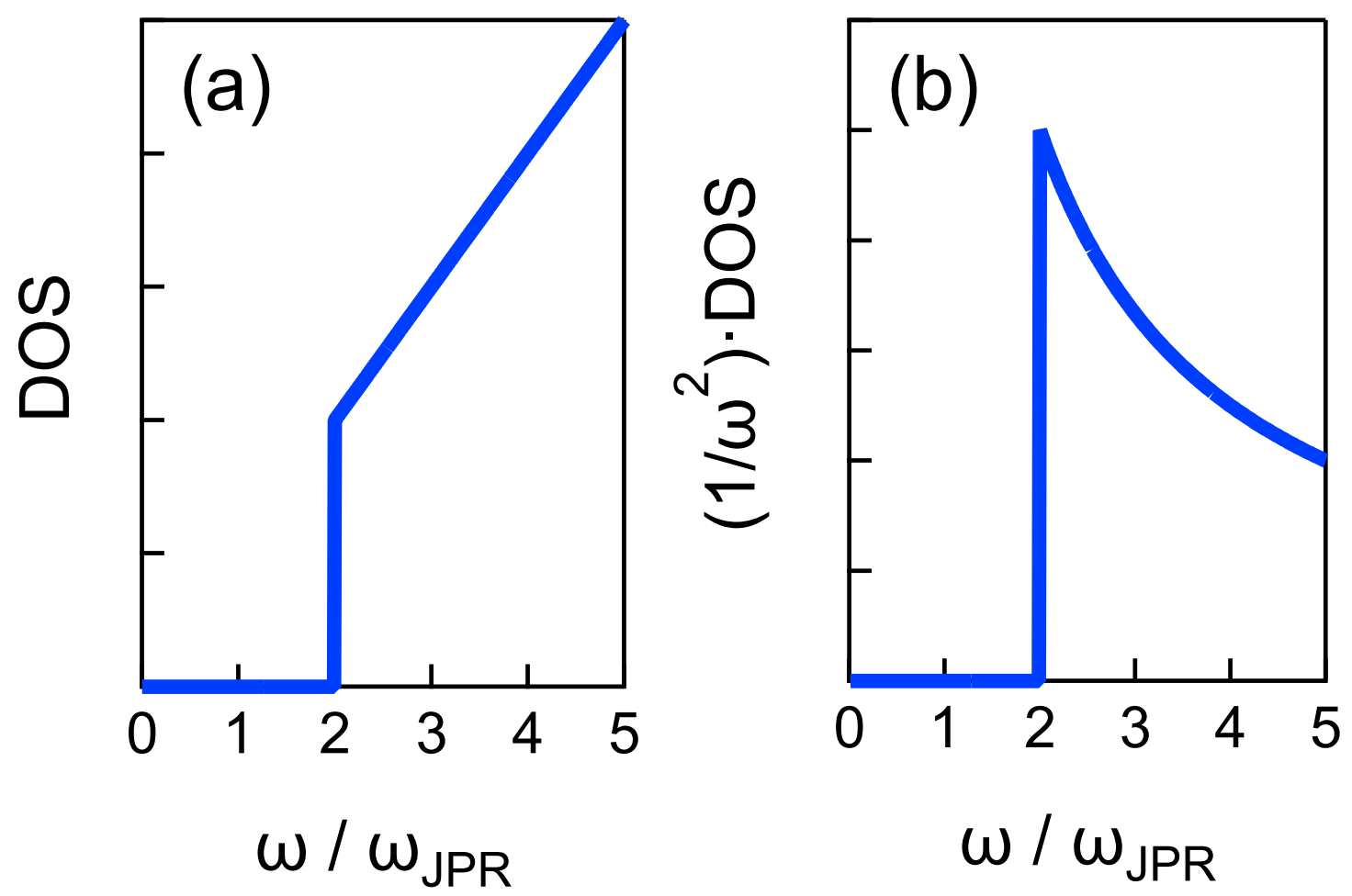

**Figure S7. (a)** Calculated "bare" bi-plasmon density of states, which grows with increasing frequency and peaks at the asymptotic value of the dispersion (not shown in the graph) for bilayer cuprates. **(b)** "Fluctuation" density of states, corresponding to the density of states in (a) multiplied by a fluctuation term. In the plot we assumed Gaussian fluctuations, which result in a $1/\omega^2$ scaling. In a real theory of YBCO, for which the transition is dominated by highly non-Gaussian BKT phase fluctuations, one expects this function to be even more peaked at zero momentum and low frequency.

In addition, we assumed that the excitation pulse at optical frequencies, whose electric field is denoted by $E_{drive}(t)$, gets absorbed by creating electron-hole pair excitations. Such driving protocol has the consequence of partially depleting the superconducting condensate and thereby suppressing the Josephson energy $\Lambda_0$ (here $\beta > 0$ is a positive constant).

Following Ref. [xx], the equations of motion for the fluctuations in the electric field and the gauge invariant phase, in the presence of an external drive and including dissipation, can be written as:

$$\partial_t D^{\theta\theta}_{\mathbf{q},k_z} = 2edD^{E\theta}_{\mathbf{q},k_z}, \tag{17}$$

$$\partial_t D^{E\theta}_{\mathbf{q},k_z} = -\frac{\Lambda_0\left(1-\beta E^2_{drive}(t)\right)+c^2(k_z)\mathbf{q}^2}{ed} D^{\theta\theta}_{\mathbf{q},k_z} - \gamma D^{E\theta}_{\mathbf{q},k_z} + 4edD^{EE}_{\mathbf{q},k_z}, \tag{18}$$

$$\partial_t D^{EE}_{\mathbf{q},k_z} = -2\gamma D^{EE}_{\mathbf{q},k_z} - \frac{\Lambda_0\left(1-\beta E^2_{drive}(t)\right)+c^2(k_z)\mathbf{q}^2}{2ed} D^{\theta E}_{\mathbf{q},k_z}, \tag{19}$$

where $D^{AB}_{\mathbf{q},k_z} = \langle A_{\mathbf{q},k_z} B_{-\mathbf{q},-k_z}\rangle$. Using the Hamiltonian in Eq. (16), we introduce the squeezing dynamics through the following equation:

$$\left(-\omega^2 - 2i\gamma\omega + \left(2\omega_{JPR}(q)\right)^2\right)\langle E^2_q(\omega)\rangle = \lambda I_{pump}\langle E^2_q\rangle_{th}. \tag{20}$$

We note that in the presence of longitudinal plasmons, the fluctuation density of states peaks at $\omega = \omega_{JPR}(q=0)$. Therefore, we expect the dominant contribution of superconducting fluctuations to the THz emission to be expressed as:

$$E^{(2)} = \int \frac{\mathrm{d}^2\mathbf{q}\mathrm{d}k_z}{(2\pi)^3} 3\alpha\lambda \frac{I_{pump}\langle E^2_q\rangle_{th}}{-\omega^2-2i\gamma\omega+4\Lambda_0(\mathbf{q},k_z)} \frac{(\Lambda_{-Q}P_{CDW,Q}+\Lambda_Q P_{CDW,-Q})(\omega^2+i\gamma\omega)}{-(\omega^2+i\gamma\omega-\Lambda_0)^2}. \tag{21}$$

Compared to the expression in Eq. (15), this "fluctuation" term introduces a new peak at $2\omega_{JPR}$, which is in agreement with the experimental observation in YBCO 6.6 (see, *e.g.*, Fig. 2 in the main text). In addition, Eq. (21) shows the same linear scaling with the drive intensity as the mean field term in Eq. (15), as observed experimentally and reported in Fig. S6.

## S6.4 Fitting formula

We now rewrite both Eq. (15) and Eq. (21) in terms of experimental observables in order to obtain the main text's expressions for $E_{CTE}^{JPR}(\omega, T)$ and $E_{CTE}^{fluct}(\omega, T)$.

We first incorporate the various pre-factors into a coefficient $A$ that expresses the relative strength of the fluctuation term to the mean field term. Note that we fitted $A$ to the YBCO 6.6 data and kept it fixed at that value for LBCO 9.5%. We then approximate $\left(\Lambda_{-Q} P_{CDW,Q} + \Lambda_Q P_{CDW,-Q}\right)$ with $\omega^2 I_{CDW}$. Here, $I_{CDW}$ is the CDW intensity, as determined experimentally with soft X-ray spectroscopy. This was taken from Ref. [v] for LBCO 9.5% and from Ref. [vi] for YBCO 6.6. The $\omega^2$ factor originates by assuming a dominant role of the dielectric background in $\Lambda_{Q,-Q}$.

We then neglect the higher order term $2\Lambda_Q \Lambda_{-Q}$ in Eq. (15) and identify $\Lambda_0$ and $\gamma$ as the experimentally determined Josephson plasma frequency squared, $\omega_{JPR}^2$, and width, $\gamma_{JPR}$, respectively. These parameters were taken from the data in Ref. [i] for LBCO 9.5% and from those of Fig. S3 of this manuscript for YBCO 6.6. We refer the reader also to Fig. S4 for a systematic analysis of $\gamma_{JPR}$ in the different compounds.

Finally, we express $\langle E^2 \rangle_{th}$ in Eq. (21) as an $N_{fluct}$ factor proportional to the fluctuating component of the superconducting order parameter, which follows the same temperature dependence as the experimentally measured second harmonic emission in YBCO 6.6 and was kept fixed for LBCO 9.5%.

Through the above procedure, we obtain the expressions introduced in the main text, which we also quote here for convenience:

$$\left|E_{CTE}^{JPR}(\omega, T)\right| = \left|\frac{I_{CDW}(T)\omega^2(\omega^2 + i\gamma_{JPR}\omega)}{-\left(\omega^2 + i\gamma_{JPR}\omega - \omega_{JPR}^2(T)\right)^2}\right| E_{opt}E_{opt}, \tag{22}$$

$$\left|E_{CTE}^{fluct}(\omega, T)\right| = \left|A\,\frac{N_{fluct}(T)\omega_{JPR}^2(T)}{\omega^2 + 2i\gamma_{JPR}\omega - 4\omega_{JPR}^2(T)}\,\frac{I_{CDW}(T)\omega^2(\omega^2 + i\gamma_{JPR}\omega)}{\left(\omega^2 + i\gamma_{JPR}\omega - \omega_{JPR}^2(T)\right)^2}\right| E_{opt}E_{opt}, \tag{23}$$

$$|E_{CTE}^{tot}(\omega, T)| = \left|E_{CTE}^{JPR}(\omega, T) + E_{CTE}^{fluct}(\omega, T)\right|. \tag{24}$$

We stress here how these expressions reproduce the experimental emission spectra in both LBCO 9.5% and YBCO 6.6 (Fig. 4 in the main text), as well as the temperature dependence of the $\omega_{JPR}$ and $2\omega_{JPR}$ peaks in both compounds (Fig. 5 in the main text).

In conclusion, our analysis shows how, similar to what has been discussed for other nonlinear optical spectroscopies[xxi,xxii,xxiii], CTE is a sensitive technique to probe fluctuations of the superconducting condensate at $T \simeq T_C$. Our approach could possibly be extended to study other kinds of phase transitions, when in conjunction with inversion or time-reversal symmetry breaking.

**REFERENCES (Supplemental Material)**

[xviii] G. Homann, M. H. Michael, J. G. Cosme, and L. Mathey, Phys. Rev. Lett. **132**, 096002 (2024).

[xix] M. H. Michael, D. De Santis, E. Demler, and P. A. Lee, *arXiv:*2410.12919 (2024).

[xx] N. Taherian, M. Först, A. Liu, M. Fechner, D. Pavicevic, A. von Hoegen, E. Rowe, Y. Liu, S. Nakata, B. Keimer, E. Demler, M. H. Michael, and A. Cavalleri, npj Quantum Materials **10**, 54 (2025).

[xxi] K. Katsumi, N. Tsuji, Y. I. Hamada, R. Matsunaga, J. Schneeloch, R. D. Zhong, G. D. Gu, H. Aoki, Y. Gallais, and R. Shimano, Phys. Rev. Lett. **120**, 117001 (2018).

[xxii] H. Chu, M.-J. Kim, K. Katsumi, S. Kovalev, R. D. Dawson, L. Schwarz, N. Yoshikawa, G. Kim, D. Putzky, Z. Z. Li, H. Raffy, S. Germanskiy, J.-C. Deinert, N. Awari, I. Ilyakov, B. Green, M. Chen, M. Bawatna, G. Cristiani, G. Logvenov, Y. Gallais, A. V. Boris, B. Keimer, A. P. Schnyder, D. Manske, M. Gensch, Z. Wang , R. Shimano, and S. Kaiser, Nat. Commun. **11**, 1793 (2020).

[xxiii] J. Yuan, L. Shi, L. Yue, B. Li, Z. Wang, S. Xu, T. Xu, Y. Wang, Z. Gan, F. Chen, Z. Lin, X. Wang, K. Jin, X. Wang, J. Luo, S. Zhang, Q. Wu, Q. Liu, T. Hu, R. Li, X. Zhou, D. Wu, T. Dong, and N. L. Wang, Sci. Adv. **10**, eadg9211 (2024).